\documentclass[fleqn,usenatbib]{mnras}

\usepackage{newtxtext,newtxmath}
\usepackage[T1]{fontenc}
\usepackage{graphicx} 
\usepackage{multirow}

\usepackage[normalem]{ulem}

\newcommand\Lya{Ly$\alpha$\,}
\newcommand\hM{h^{-1} M_\odot}

\newcommand\mG{\ensuremath{\delta_m}}

\newcommand\dfG{\ensuremath{\delta_{F}}}
\newcommand\dfWF{\ensuremath{\delta_F^{\mathrm{WF}}}}

\newcommand\ic[1]{\iffalse #1 \fi}
\newcommand\kmsMpc{\mathrm{km\,s^{-1}\,Mpc^{-1}}}
\newcommand{\hMpc}{\ensuremath{h^{-1}\,\mathrm{cMpc}}}
\newcommand{\cxc}{CLAMATO\ensuremath{\times}COSTCO}
\newcommand{\hl}{\mathrm{halo}}
\newcommand{\Fuvb}{\ensuremath{A_{\mathrm{UVB}}}}
\newcommand{\tauraw}{\ensuremath{\tau_{\mathrm{raw}}}}

\newcommand{\sanbai}{\texttt{The Three Hundred}}

\title[AGN Feedback Effect in Protoclusters]{The Effect of AGN Feedback on the Lyman-$\alpha$ Forest Signature of Galaxy Protoclusters at $z\sim 2.3$}

\author[C.Z. Dong et al.]{
Chenze Dong,$^{1,2}$\thanks{E-mail: dong-chenze@g.ecc.u-tokyo.ac.jp}
Khee-Gan Lee,$^{1,2}$
Weiguang Cui,$^{3,4}$
Romeel Davé$^{3,5}$
and Daniele Sorini$^{6}$
\\
$^{1}$Kavli Institute for the Physics and Mathematics of the Universe (WPI), UTIAS, The University of Tokyo, Kashiwa, Chiba 277-8583, Japan\\
$^{2}$Center for Data-Driven Discovery, Kavli IPMU (WPI), UTIAS, The University of Tokyo, Kashiwa, Chiba 277-8583, Japan\\
$^{3}$Institute for Astronomy, University of Edinburgh, Royal Observatory, Blackford Hill, Edinburgh EH9 3HJ, UK\\
$^{4}$Departamento de Física Teórica and CIAFF, Modulo 8 Universidad Autónoma de Madrid, 28049 Madrid, Spain\\
$^{5}$University of the Western Cape, Bellville, Cape Town 7535, South Africa\\
$^{6}$Institute for Computational Cosmology, Durham University, South Park Road, DH1 3LE, United Kingdom\\
\\
}

\date{Accepted XXX. Received YYY; in original form ZZZ}

\pubyear{2024}

\begin{document}

\label{firstpage}
\pagerange{\pageref{firstpage}--\pageref{lastpage}}
\maketitle
\begin{abstract}
The intergalactic medium (IGM) in the vicinity of galaxy protoclusters are interesting testbeds to study complex baryonic effects such as gravitational shocks and feedback. 
Here, we utilize hydrodynamical simulations from the SIMBA and The Three Hundred suites to study the mechanisms influencing large-scale Lyman-$\alpha$ transmission in $2<z<2.5$ protoclusters.
We focus on the matter overdensity-Lyman-$\alpha$ transmission relation $(\delta_m-\delta_F)$ on Megaparsec-scales in these protoclusters, which is hypothesized to be sensitive to the feedback implementations.
The lower-density regions represented by the SIMBA-100 cosmological volume trace the power-law $\delta_m-\delta_F$ relationship often known as the fluctuating Gunn-Peterson approximation.
This trend is continued into higher-density regions covered by simulations that implement stellar feedback only.
Simulations with AGN thermal and AGN jet feedback , however, exhibit progressively more Lyman-$\alpha$ transmission at fixed matter overdensity.
Compared with the 7 protoclusters observed in the COSMOS field, only 2 display the excess absorption expected from protoclusters.
The others exhibit deviations: 4 show some increased transparency suggested by AGN X-ray thermal feedback models while the highly transparent COSTCO-I protocluster appears to reflect intense jet feedback.
Discrepancies with the stellar-feedback-only model suggests processes at play beyond gravitational heating and/or stellar feedback as the cause of the protocluster transparencies. 
Some form of AGN feedback is likely at play in the observed protoclusters, and possibly long-ranged AGN jets in the case of COSTCO-I.
While more detailed and resolved simulations are required to move forward, our findings open new avenues for probing AGN feedback at Cosmic Noon.
\end{abstract}

%%  We find, SIMBA model forecasts that all protoclusters exceeding a specific mass will exhibit considerable large-scale gas heating, but it is mainly attributed to the effect of environment, where the highly-collimated jet employed in SIMBA has a negligible influence. The Nyx run without feedback falsifies the shock heating scenario. 

\section{Introduction} \label{sec:intro}
Galaxy clusters did not form in a day; these giants were still in their adolescence stage at redshift $z>2$, in structures commonly known as ``galaxy protoclusters". At this cosmic noon epoch, they were still in an active stage of mass assembly, and their member galaxies may or may not end up in a present-day cluster \citep{2013PC_obs}. 
Protoclusters contribute a significant fraction to the cosmic star formation rate density \citep{2017PC_SF} at Cosmic Noon when cosmic star-formation activity was at its peak.
Galaxy protoclusters do not necessarily exhibit very conspicuous galaxy overdensities, with most nascent protocluster cores accumulating masses of $M \lesssim 10^{14}\, M_\odot$ by this epoch, 
and the mass growth of these protoclusters is not necessarily monotonic (i.e.\ the most massive protocluster cores at $z\sim 2$ will not always grow into the most massive $z\sim 0$ clusters, for example, \citealt{remus:2023, Cui2020}). 

Aside from their important role in star formation and massive galaxy assembly, protoclusters also show interesting interactions with the intergalactic medium (IGM), \citep[for example][]{kikuta:2019, 2022Preheat_sim, emonts:2023, cai2017}.
After the end of cosmic reionization around $z\sim 6$, the free streaming photons and the expansion of the universe quickly erase the thermal memory of IGM toward reionization. The inverse Compton scattering of electrons off the Cosmic Microwave Background (CMB) is also an important cooling mechanism at high redshift, but its cooling efficiency is about one magnitude lower than the cooling caused by cosmic expansion at $2<z<3$ \citep{McQuinn:2016}.
The equilibrium between adiabatic expansion and photo-ionisation heating leads to a power-law relation between the temperature and density in the IGM,  $T = T_0 (1+\delta)^{\gamma - 1}$ \citep{1997TDR}, where $\delta=\rho/\bar\rho - 1$ is the overdensity, $T_0$ is the typical temperature for IGM with an average density.  The temperature-density equation-of-state index, $\gamma\approx 1.7$ \citep{2015fluxPDF, 2018Hiss, 2018Rorai}, is a parameter arising from the balance between recombination and photo-ionisation. 
This leads to a relation between hydrogen Lyman-alpha (Ly$\alpha$) forest optical depth and IGM state, often known as the Fluctuating Gunn-Peterson Approximation \citep[FGPA;][]{1998FGPA, 2002FGPA}:
\begin{equation} \label{eq:FGPA}
    \tau_{\mathrm{Ly}\alpha} \propto \frac{T^{-0.7}}{\Gamma_{UV}}(1+\delta)^2 \propto (1+\delta)^\beta,
\end{equation}
where $\Gamma_{UV}$ corresponds to the background ultraviolet (UV) photo-ionisation rate, and $\beta = 2 - 0.7(\gamma - 1)$ is another positive power-law index. 
By observing Ly$\alpha$ absorption, this relation facilitates the probing of overdensity $\delta$ along lines of sight to background objects, for example, CLAMATO (\citealt{2014FirstLya}, \citealt{2022CLAMATO},  \citealt{2018CLAMATO}) and LATIS (\citealt{2020LATIS}).
Protoclusters are usually the sites with the highest overdensities and strongest \Lya{} absorption, so this relation further enables the detection of protoclusters at Cosmic Noon \citep{2015Tomopc, 2016PCLya, cai:2016, 2021PCLya,2022LATIS_PC}. 

Baryonic processes associated with galaxy formation in and around the protoclusters add great complexity to the connection between the IGM and galaxy protoclusters. 
Stellar and AGN feedback can inject energy into the IGM as they regulate the star formation, potentially causing deviations from the FGPA due to collisional heating and ionization. Furthermore, even shock heating purely driven by gravitational collapse can raise the temperature of the gas within high-mass halos \citep{2005Gal_gas_accretion}, eventually driving the evolution into the complex and ``missing'' phases of the IGM at late times \citep{dave:2001, 2006MissBaryon}.
Early works \citep{2001Lya_FB_obs} hinted at this more complex connection by inspecting individual quasar spectra.
On the simulation side, various studies assessed the impact of feedback processes both on the summary statistics of the \Lya forest \citep[for example][]{Kollmeier_2003, Christiansen:2020, 2021Lya_FB, Burkhart_2022,  Tillman_2023, Tillman:2023, Khaire:2023, Khaire:2024} and on \Lya absorption in the circumgalactic medium of galaxies and quasars \citep{Kollmeier_2006, Meiksin_2015, Meiksin_2017, Turner_2017, Sorini_2018, Sorini_2020, Appleby_2021, Mallik:2023, Maitra:2024}.
\citet{2022Preheat_sim} also predicted an elevation of IGM temperature by the manual phenomenological implementation of AGN feedback.

Recently, \citet{2023Preheat} found a particularly interesting case, the COSTCO-I protocluster, which potentially holds strong implications for the interplay between heating processes on the Ly$\alpha$ forest in protoclusters. 
This protocluster was first identified in density reconstructions of the COSMOS spectroscopic catalogue \citep{ata2021} and resulting constrained simulations (COSTCO project; \citealt{2022COSTCO}) as the progenitor of a late-time $M = (4.6 \pm 2.2) \times 10^{14}\,M_\odot$ galaxy cluster. 
However, a comparison with the Wiener-filtered tomographic IGM map from the CLAMATO survey \citep{2022CLAMATO} indicated that COSTCO-I is anomalously transparent to Lyman-$\alpha$ absorption, contrary to the significant absorption expected through FGPA given its galaxy overdensity. 
It is a special case among protoclusters residing in the overlapping volume of CLAMATO and COSTCO (hereafter \cxc{} protoclusters), as none of the other protoclusters have such transparent Ly$\alpha$ absorption at first glance.
Moreover, the unexpectedly high transmission is still evident after averaging with a $r=15$ cMpc/h tophat kernel, indicating that this phenomenon spans multiple physical Megaparsecs. 
Recent work by \citet{miller2022} suggests that the UV radiation from protocluster AGN seems unlikely to be the cause of extended Ly-$\alpha$ transmission to the extent that we see unless an ultra-luminous quasar is present \citep{schmidt:2018}. 
Therefore, according to Equation~\ref{eq:FGPA}, such a Ly$\alpha$ transparency could be attributed to a high IGM temperature across several physical Mpc, given the uniform UV background and overdensity confirmed by COSTCO. 
We hypothesise that the extended gas around COSTCO-I is a possible site of large-scale heating processes other than UV background ionisation, which breaks the power-law relation between the temperature and density of IGM. 
However, the mechanism that caused the heating still remains unclear, as it could be some combination of AGN and/or stellar feedback along with gravitational shock heating.

In this paper, we use the SIMBA cosmological galaxy formation simulation and The Three Hundred Cluster zoom simulation suite to investigate the impact of gravitational shock heating and various feedback processes on the \Lya\ forest within and around large halos at $z=2.5$. In Section \ref{section:data}, we introduce the data adopted in this study and describe how we derive the quantities of interest. In Section \ref{section:result}, we present our main findings based on the measurement from both observations and simulations. In Section \ref{section:discussion}, we discuss the results and summarise the study. Throughout this paper, we adopt the cosmology of $H_0 = 100 h\ \kmsMpc= 70\ \kmsMpc$, $\Omega_m = 0.315$, and $\Omega_\Lambda = 0.685$ for the observational data to maintain consistency with the cosmology in \citet{2023Preheat}.

\section{Data} \label{section:data}
In order to quantify the heating of IGM around massive protoclusters in simulations and compare with the observed case with large-scale gas heating, we focus on the following quantities: overdensity \mG and Lyman-alpha transmission \dfG. 
The definition and measurement methodology of these quantities for the \cxc{} observational data are described in Section \ref{ss:measure_obs}. The simulation suites used in this study are introduced in Section \ref{ss:sim}, and in Section \ref{ss:measure_sim} we applied the same measurement to the simulations.

\subsection{CLAMATO and COSTCO Observations} \label{ss:measure_obs}
This paper was motivated by the observational discovery of the COSTCO-I galaxy protocluster at $z=2.30$ \citep{2016PCLya, 2022COSTCO, 2023Preheat}.
The unusual nature of this protocluster was first hinted at by \citet{2016PCLya}, but the official discovery was made through constrained N-body simulations \citep{2022COSTCO} based on the reconstructed matter density field of the COSMOS galaxy spectroscopic sample in the redshift range $2.0<z<2.5$ \citep{ata2021}. 
To briefly summarise, the initial conditions (ICs) of the N-body simulations were first sampled using the \texttt{COSMIC-BIRTH} \citep{2020COSMIC_BIRTH} algorithm, conditional to the observed spectroscopic galaxy redshift distribution at $2.00 < z_{\mathrm{obs}} < 2.52$ in the COSMOS field.
\citet{2022COSTCO} then ran the \texttt{PKDGRAV3} N-body code \citep{2017PKDGRAV3} on the ICs derived from this analysis, with a total of 57 posterior realisations and produced simulation snapshots at $z=2.3$ as well as at $z=0$.
This allows us to compute various quantities for all the \cxc{} protoclusters, including some that are not typically accessible through observations such as the matter density field.
In addition, this allows us to estimate the uncertainties related to the derived quantities thanks to the posterior realisations.

Table \ref{tab:pc_info} lists all the \cxc{} protoclusters that we analyse in this study. 
Among the entries, ZFIRE was first detect through medium-band imaging \citep{Spitler:2012} and later confirmed by Keck / MOSFIRE spectroscopy \citep{Yuan:2014}. 
Hyperion \citep{Diener:2015, Chiang:2015, Casey:2015, 2016PCLya, Wang:2016} is a complex of galaxy protoclusters identified at redshift $2.423<z<2.507$, where only Hyperion 1, 3, 4 and 5 are enclosed in the volume of CLAMATO.
COSTCO-I and COSTCO-III are newly discovered galaxy protocluster reported by \citet{2022COSTCO} based on constrained-simulation analysis. 

\begin{table}
    \centering
    \scalebox{1}{
    \centering
        \begin{tabular}{ccccc}
        Protocluster  & RA & Dec & $z_{\mathrm{obs}}$ & Final Mass ($\hM$) \\ \hline
        ZFIRE & 150.094 & 2.251 & 2.095 & $(1.2 \pm 0.3) \times 10^{15}$ \\
        Hyperion 1 & 150.093 & 2.404 & 2.468 & \multirow{4}{*}{$(2.5 \pm 0.5)\times 10^{15}$} \\
        Hyperion 3 & 149.999 & 2.253 & 2.444 &  \\
        Hyperion 4 & 150.255 & 2.342 & 2.469 &  \\
        Hyperion 5 & 150.229 & 2.338 & 2.507 &  \\
        COSTCO-I & 150.11 & 2.161 & 2.298 & $(4.6 \pm 2.2) \times 10^{14}$ \\
        COSTCO-III & 150.129 & 2.275 & 2.16 & $(5.3\pm 2.6)\times 10^{14}$
        \end{tabular}
    }
    \caption{\cxc{} galaxy protoclusters analysed in this work. Note that \citet{2022COSTCO} only reported the final mass for Hyperion clusters as the sum of all member clusters (Hyperion 1-7), including Hyperion 2; Hyperion 6 and Hyperion 7 are outside the volume of the CLAMATO Lyman-$\alpha$ map.}
    \label{tab:pc_info}
\end{table}

To derive the quantities of interest, we first measured the matter overdensity, \mG{} around the protoclusters from the N-body outputs.
To do this, we converted the particle data from the simulation output into a grid of $\delta_{m, \mathrm{raw}}$ and used the mean density of the universe to normalize the matter density,
\begin{equation}
    \delta_{m, \mathrm{raw}} = \rho / \bar{\rho} - 1
\end{equation}
Then, we smoothed the $\delta_{m, \mathrm{raw}}$ field with a Gaussian kernel of 7.5 cMpc/h smoothing radius, and took the value of \mG{} at the positions of protoclusters as the overdensity associated with the protocluster.

We measured the Ly$\alpha$ forest transmission fluctuation, $\delta_F = F/\bar{F}-1$ (where $F = \exp(-\tau_{\mathrm{Ly}\alpha})$ is the Ly$\alpha$ transmission, and $\bar{F}$ is the mean Ly$\alpha$ transmission at $z=2.5$ given by \citealt{Becker:2013}) associated with the protoclusters based on the Wiener-filtered \dfWF{} from the CLAMATO survey \citep{2022CLAMATO}. To match the measurement of overdensity above, we simply smoothed the \dfWF{} field with a Gaussian kernel of 7.5 cMpc / h and evaluated \dfG{} at the position of the protocluster, which was used to compare with the \dfG{} matter overdensity. In addition, Wiener filtering under irregular sightline sampling can lead to biased reconstructions, $\delta_F$, compared to the underlying truth (see \ref{ap:obs_cali}). 
We corrected the bias by introducing a rescaling factor for the \dfWF{} values of the \cxc protoclusters, calibrated from simulated mock sightlines.

\subsection{Simulations} \label{ss:sim}
In order to explore a wide range of halo masses, larger simulation boxes are generally desired. 
However, this typically requires a trade-off between the simulation scale and the resolution given finite computational resources. 
In our work, we aim to achieve adequate resolution while retaining the ability to study massive halos ($M_{\hl} > 10^{14} \hM$ at $z=0$) when picking the runs in each simulation suite. 
We, therefore, use the large volume available in the \texttt{SIMBA} suite of cosmological hydrodynamical simulation, i.e.\ the $L=100\,\hMpc$ box. 
We also study \sanbai{}, a series of zoom-in simulations that focus on the evolution of the most massive galaxy clusters at $z=0$.
These simulations have been run with several different feedback prescriptions.
We introduce the simulation suites below with a focus on their jet feedback implementation and summarize them in Table \ref{tab:sims}.
Due to the limited snapshots available in \sanbai{}, we study the property of massive protoclusters with snapshots at $z=2.5$, which correspond to the highest redshift in the CLAMATO volume. 
The choice of snapshot redshift can be justified for the following reasons. 
Firstly, the median redshift of \cxc{} protoclusters is $z=2.444$ (see Table \ref{tab:pc_info}), therefore adopting $z=2.5$ is nominally the closest snapshot. 
Secondly, most clusters in \sanbai{} sample have larger final masses ($>10^{15} \hM$) than \cxc{} sample, thus a slightly higher redshift for the simulated sample facilitates a better match with the evolutionary stage of the observed protoclusters.

\subsubsection{SIMBA}\label{ssec:SIMBA}
The \texttt{SIMBA} simulation \citep{2019SIMBA} is a cosmological simulation suite based on a forked version of the \texttt{Gizmo} \citep{2015GIZMO} Gravity+hydrodynamics solver. The meshless-finite-mass (MFM) scheme in \texttt{Gizmo} gets rid of artificial viscosity in its hydrodynamic implementation, which handles shocks in a more precise and realistic way. The \texttt{SIMBA} simulation includes a variety of input physics.
The radiative cooling and photo-ionisation heating are implemented using \texttt{Grackle-3.1} \citep{Smith:2017} with an on-the-fly, and self-consistent self-shielding treatment \citep{Rahmati:2013}. \texttt{SIMBA} adopts the ionizing background in \citet{Haardt:2012} (HM12), which is further modified to account for self-shielding. 
The star formation formalism is similar to its precursor, \texttt{MUFASA} \citep{2016MUFASA}. 
The star formation rate (SFR) is calculated with an $H_2$-based \citet{Schmidt:1959} law, where the SFR is proportional to the $H_2$ density and the inverse of dynamic time.
\texttt{SIMBA} incorporates the stellar feedback from Type Ia \& Type II Supernovae (SNe) and Asymptotic Giant Branch (AGB) stars. The \texttt{SIMBA} applied the decoupled two-phase winds for stellar feedback, of which the effect is aligned with high-resolution FIRE simulation. (Feedback In Realistic Environments, \citealt{Hopkins:2014})

The major improvement in \texttt{SIMBA} sub-grid models relative to \texttt{MUFASA} is the black hole physics. 
For the black hole accretion, \texttt{SIMBA} employs a combination of Bondi-Hoyle-Lyttleton accretion \citet{Bondi:1952} (from hot gas) and Torque-limited accretion \citet{Hopkins:2011} (from cold gas).
The AGN feedback is modelled in a two-phase approach, featuring criteria on the black hole mass $M_{\mathrm{BH}}$ and Eddington ratio $f_{\mathrm{Edd}}$. 
The  ``radiative" mode is activated for the black holes in high accretion state ($f_{\mathrm{Edd}} > 0.2$) or with low mass ($M_{\mathrm{BH}}<10^{7.5}M_{\odot}$), in which the black holes exert kinetic kick on surrounding gas to create  ``AGN winds" with a speed of $\sim 10^3\,\mathrm{km/s}$. For more massive black holes with lower accretion rate, the feedback enters ``jet" mode, where the gas at black hole vicinity is heated to $\gtrsim 10^7\,\mathrm{K}$ and launched with a speed of $\sim 7000\,\mathrm{km/s}$.
The crucial feature of SIMBA's AGN feedback arises from the directionality of momentum ejection. 
For both wind and jet feedback, the ejected gas launches from the direction perpendicular to the inner disc around the SMBH with an opening angle of zero. Given the high velocity of the jet feedback, it leads to a highly collimated bipolar outflow, which has an impact on the scale of observed radio lobes. 
The authors argue the highly collimated implementation of feedback is later alleviated by other processes on longer time scales, such as the precession of the inner disk and the interaction of outflow and ambient gas.
For more detail, we refer the readers to the introductory paper of SIMBA \citep{2019SIMBA}.

\texttt{SIMBA} has simulation boxes with settings different in different box sizes and resolutions publicly available.
In this study, We adopt the flagship run (hereafter SIMBA-100) with the box size of 100 cMpc/h and $2\times1024^3$ particles to increase the volume and hence halo mass coverage compared with the 50 \hMpc{} versions.

\subsubsection{The THREE HUNDRED}
\sanbai{} project \citep{2018the300_Gadget, 2022the300_SIMBA} is a set of re-simulations of 324 large galaxy clusters. 
These clusters were first selected from the 1 cGpc/h MDPL2 dark-matter-only simulation \citep{2016MDPL2} based on their halo mass at $z=0$ ($M_{h, z=0} > 6.42\times 10^{14} M_\odot$). 
A spherical volume with a radius $r=15$ cMpc/h  was then selected as the re-simulation region. 
The ICs were generated by the \texttt{GINNUNGAGAP} code with the re-simulation region split into gas and dark matter particles, and the outer regions are resampled using low-resolution particles to save the computation cost. 
\sanbai{} resimulation set provides a mass-complete sample of the extremely massive halos with flexibility in baryon model prescriptions. 
Our study includes the runs with GadgetX, GadgetMUSIC and GIZMO-SIMBA models for all 324 clusters.

The detail of three prescriptions in \sanbai{} is listed as follows.
\begin{itemize}
    \item \texttt{GadgetMUSIC} (hereafter 300-GadgetMUSIC): 
    This prescription is based on \citet{Sembolini:2012}, running with \texttt{GADGET-3} Tree-PM code, an updated version of \texttt{GADGET-2} \citep{Springel:2005} 
    The classic entropy-conserving SPH formulation with a 40 neighbour M3 interpolation kernel is used in the hydrodynamic solver of 300-GadgetMUSIC.
    It adopts a metal-independent approach for radiative cooling and a homogeneous UV background (UVB) by \citet{Haardt:2001} (HM01) for photo-ionisation heating. 
    On star formation, it follows the hybrid multi-phase model in \citet{Springel:2003} with the \citet{Salpeter:1955} initial mass function (IMF). 
    The kinetic stellar feedback and thermal stellar feedback in 300-GadgetMUSIC are realized based on \citet{Springel:2003} and the 2-phase model in \citet{Yepes:1997}, respectively. 
    There is neither black hole accretion nor black hole feedback implemented in 300-GadgetMUSIC.
    
    \item \texttt{GadgetX} (hereafter 300-GadgetX): 
    This prescription has a few differences compared with 300-GadgetMUSIC. 
    It uses an updated SPH scheme \citep{Beck:2015} with improvement in artificial thermal diffusion, time-dependent artificial viscosity, high-order Wendland C4 interpolating kernel and wake-up scheme. 
    On the gas heating and cooling, 300-GadgetX adopts \citet{Haardt:1996} (HM96) UVB and a metal-dependent cooling calculation \citet{Wiersma:2009}.
    It models star formation using \citet{Tornatore:2007} and \citet{Chabrier:2003} IMF. 300-GadgetX also implemented kinetic stellar feedback following \citet{Springel:2003}.
    Contrary to 300-GadgetMUSIC, 300-GadgetX suite has a black hole physics implementation by \citet{Steinborn:2015}, which includes a multi-phase gas accretion and thermal AGN feedback.
    
    \item \texttt{GIZMO-SIMBA} (hereafter 300-SIMBA): 
    This prescription adopted a slightly modified model based on the SIMBA cosmological runs. 
    Because the resolution of The Three Hundred is downgraded compared to SIMBA-100, \citet{2022the300_SIMBA} re-calibrated the feedback model based on a series of criteria related to stellar mass and halo mass in their Section 3.2. 
    The modification is mainly motivated by the deficit of star formation in satellite galaxies. The star formation efficiency was then tuned higher to alleviate the deficit, which led to an overproduction of stars in central galaxies.
    In turn, the jet feedback was tuned more powerful (with a max speed of $\sim15000\,\mathrm{km/s}$) to regulate stellar mass in the central galaxies. 
    As a result, the jets in 300-SIMBA are the strongest among the simulation suites in this study, which results in a relatively flat gas density profile at z=0 \citep{Li2023}. 
\end{itemize}

Apart from the snapshot at $z=2.5$, we also use the output of halo finder \texttt{AHF} \citep{Knollmann:2009} to determine the position of the most massive halos in each re-simulation.

\begin{table*}
    \centering
    \scalebox{1}{
    \centering
        \begin{tabular}{ccccc}
Simulation & Box size (cMpc/h) & Cosmology ($h, \sigma_8, \Omega_m, \Omega_b$) & Mass resolution ($\hM$) & Feedback model \\ \hline
%\multicolumn{1}{l}{SIMBA} &  &  &  &  \\ \hline
SIMBA-100 & 100 & ($0.68, 0.82, 0.3, 0.048$) & $6.51 \times 10^7$ & SIMBA full physics \\
%\multicolumn{1}{l}{THE-300 (1 cGpc/h parent box)} &  &  &  &  \\ \hline
300-SIMBA & zoom-in & ($0.678, 0.823, 0.307, 0.048$) & $1.27 \times 10^9$ & SIMBA full physics\\
300-GadgetX & zoom-in & ($0.678, 0.823, 0.307, 0.048$) & $1.27 \times 10^9$ & Stellar \& AGN thermal feedback \\
300-GadgetMUSIC & zoom-in & ($0.678, 0.823, 0.307, 0.048$) & $1.27 \times 10^9$ & Stellar feedback
    \end{tabular}}
    \caption{Hydrodynamical simulations used in this work. We specify the alias, box size, cosmology, resolution and feedback model in each run.}
    \label{tab:sims}
\end{table*}

\subsection{Mock Lyman-alpha Skewers} \label{ss:mock_skewer}
As all the simulations used in this study utilized \texttt{GADGET} as their backbone, we use a light-weighted Python package \href{https://bitbucket.org/broett/pygad/src/master/}{pygad} to handle the simulation output. 
The package integrates the modules for reading simulation data with automatically inferred units, producing maps of quantities with the given SPH kernel, and generating the Lyman-$\alpha$ optical depth. 
When generating the Lyman-$\alpha$ skewers, the package will use the cosmology defined in the simulation output (see Table \ref{tab:sims}). The cosmologies adopted in simulations have subtle percent-level differences from the one used for the observational data reduction \citep{2022CLAMATO,2022COSTCO}. 
In this paper, we are aiming at a largely qualitative comparison, so we do not consider these differences to be important. 
Future analyses should, however, take careful account of the assumed cosmology.

In addition, the cosmological simulations suites adopted in this study adopted different empirical UVBs, which may cause inconsistencies between the simulation and observed Lyman-$\alpha$ mean flux. To facilitate a fair comparison between simulations and observation,s we calibrated the transmission flux in simulations following the approach in \citet{Christiansen:2020}. The details can be found in Appendix \ref{ap:sim_cali}.

\subsection{Simulation Observables} \label{ss:measure_sim}
For the simulations, we first identified the positions of massive halos ($M_h > 10^{14} M_\odot$ at present day) in SIMBA-100 and the most massive halo in each of the \sanbai{} zoom-in simulation at $z=2.5$,
then calculated the same Lyman-$\alpha$ forest and mass proxies as for \cxc{} protoclusters: \mG{} and \dfG{}.
This involved computing the gridded matter density and Lyman-$\alpha$ transmission for each zoom-in simulation snapshot using \texttt{pygad}. 
We then applied the same smoothing kernels as described in \ref{ss:measure_obs}, and finally evaluated the value at the position of corresponding halos. Some of the central halos in \sanbai{} were at the edge of the simulation zoom-in regions and were not properly simulated at full resolution, so we removed these halos from our subsequent analysis.

\section{Results} \label{section:result}
\begin{figure*}
    \centering
    \scalebox{0.4}{\includegraphics{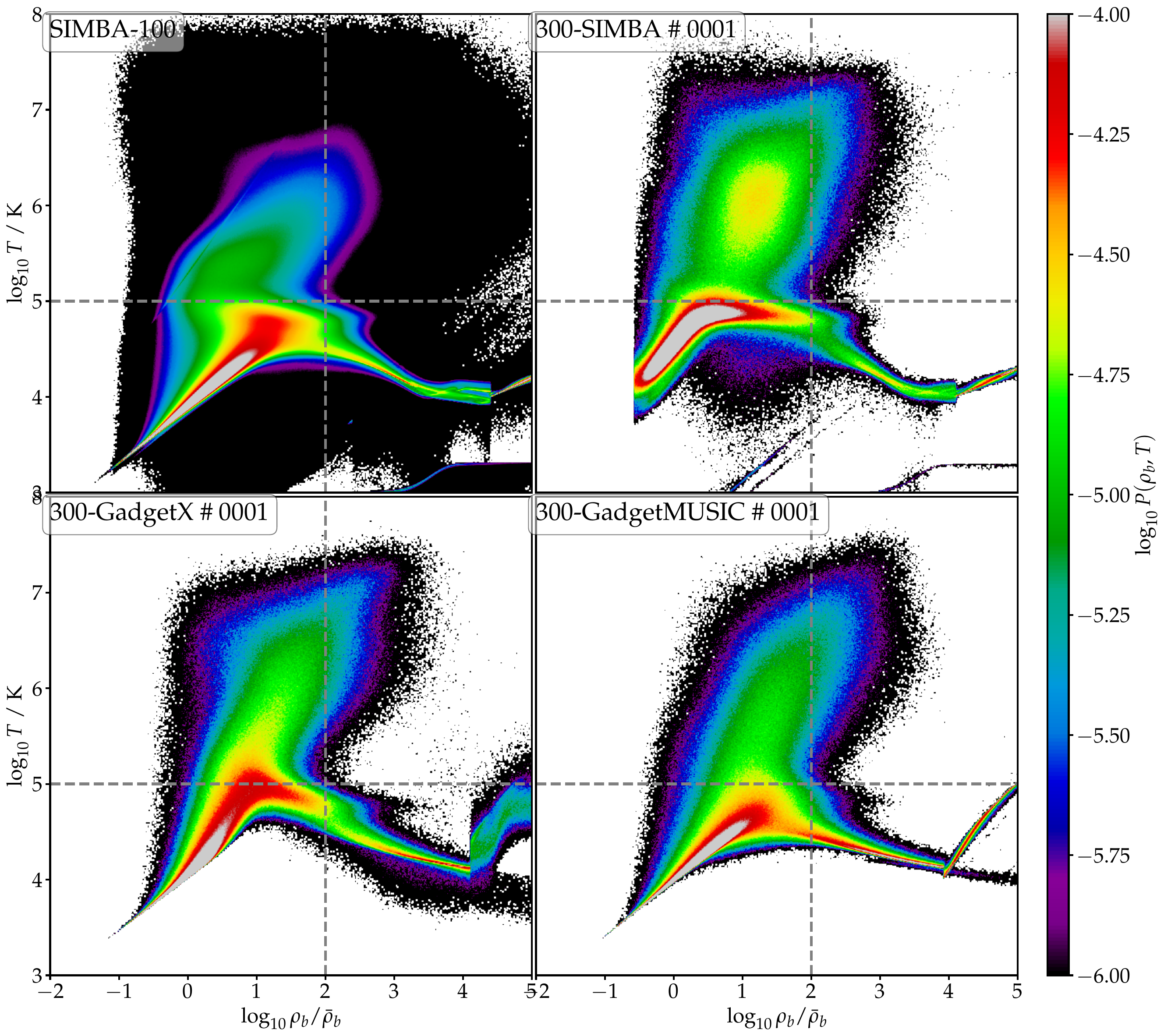}}
    \caption{The mass-weighted temperature-density relation for SIMBA-100 (upper left) and three feedback prescriptions (rest panels) on \sanbai{} cluster \#0001 at $z=2.5$. 
    In each panel, the vertical grey dash line represents $\log_{10}\rho_b/\bar{\rho}_b = 2$ while the horizontal line is $T = 10^5 \mathrm{K}$. 
    The two dash lines divide the $\rho-T$ phase space into diffused cold gas (lower left), warm-hot intergalactic medium (WHIM, upper left), hot halo gas (upper right) and condensed gas (lower right). 
    Among the \sanbai{} runs, the fraction of the WHIM component in each panel reflects the strength of feedback.}
    \label{fig:TDR_compare}
\end{figure*}

Figure~\ref{fig:TDR_compare} illustrates the distribution of gas particles on the temperature-density phase diagram. 
As a "baseline" run for this study, SIMBA-100 reproduces the power-law temperature-density relation in the diffuse cold gas regime, and shows a population of WHIM gas on $T>10^5$ K. 
All \sanbai{} phase diagrams exhibit a more notable WHIM component since \sanbai{} runs are zoom-in simulations around extremely massive regions rather than a regular cosmological volume used in SIMBA-100. 
Among these phase diagrams, the WHIM fraction of 300-SIMBA run exceeds 300-GadgetX run, and 300-GadgetX outmatches 300-GadgetMUSIC run, which reflects the latter has relatively weak feedback. 

\begin{figure*}
    \centering
    \scalebox{0.4}{\includegraphics{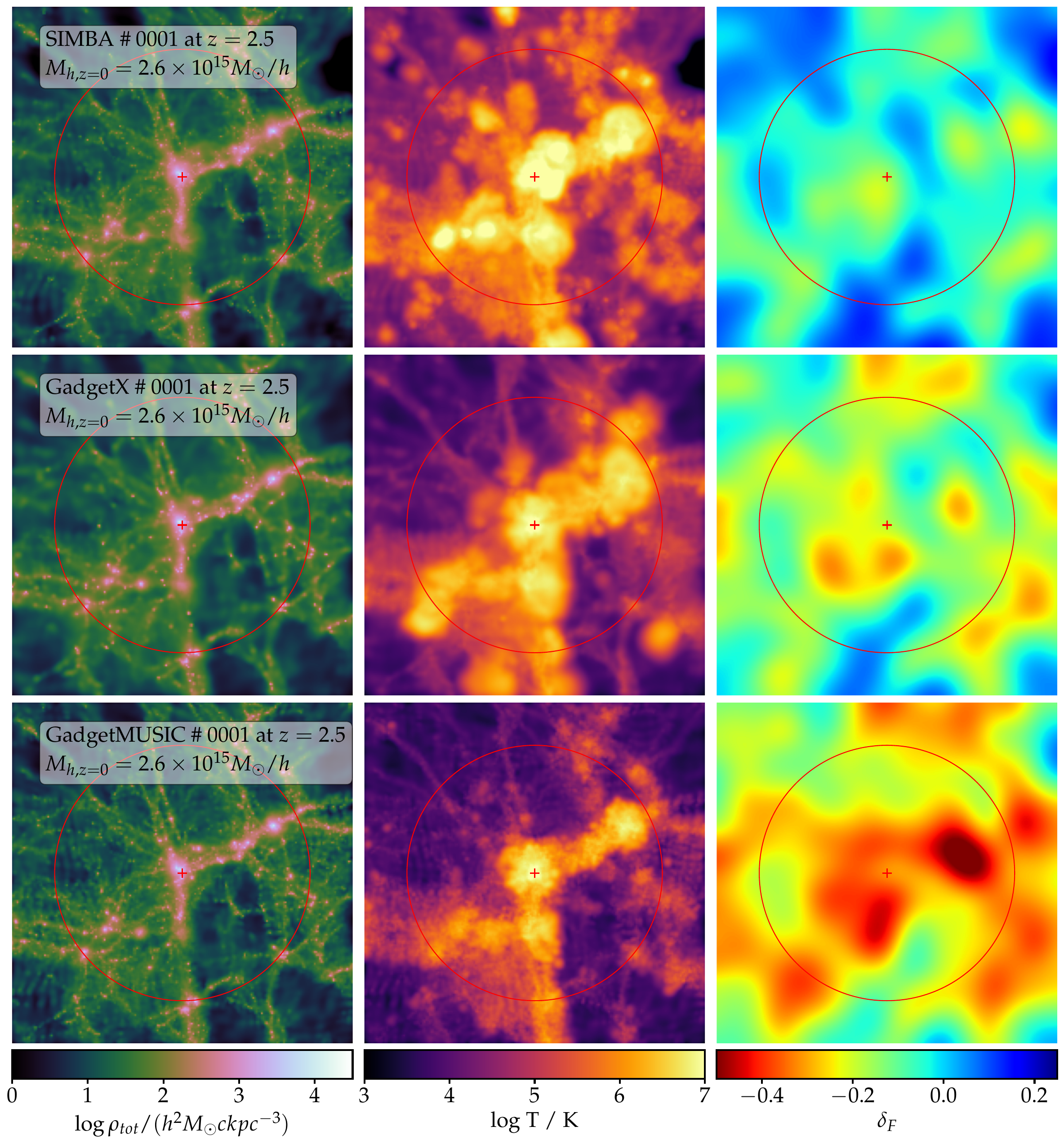}}
    \caption{$20 \times 20 \times 2$ cMpc/h slice of The Three Hundred \# 0001. 
    The left column shows matter density maps, the middle panels are the gas temperature maps, and the right panel are the maps of Lyman-$\alpha$ transmission fluctuation $\delta_F$. 
    The $\delta_F$ map is smoothed with a Gaussian kernel with $R=1$ cMpc/h. From top to bottom, the maps are from GIZMO-SIMBA, GadgetX and GadgetMUSIC prescriptions, respectively. 
    The centres of halos are marked with red plus signs; the red circles illustrate the $7.5$ cMpc/h scale used in Section \ref{ss:measure_obs} and Section \ref{ss:measure_sim}. 
    The jet feedback of the 300-SIMBA run significantly heats up the surrounding IGM, leading to a high overall Lyman-$\alpha$ transmission even with the massive galaxy protocluster overdensity.}
    \label{fig:showcase_halos}
\end{figure*}

In Figure \ref{fig:showcase_halos} we display the projected matter density, gas temperature, and Lyman-$\alpha$ transmission maps of The Three Hundred cluster \#0001. 
The matter density distribution is almost identical among the different feedback prescriptions, 
which supports the idea that even strong feedback does not significantly alter the overall matter distribution \citep{2023Baryon_fb_cosmicweb}.
The gas temperature maps, however, show distinct differences. The 300-SIMBA AGN jet-feedback runs show the highest overall IGM temperature in \sanbai{}, which is centered on hot ($> 10^7$ K) bubbles at the nodes of the cosmic web and elevated gas temperature ($\sim 10^5$ K) that extends over several Megaparsecs into regions with milder overdensity. 
As a result, the Lyman-$\alpha$ forest absorption is almost completely suppressed in 
this 300-SIMBA protocluster (top-right panel of \ref{fig:showcase_halos}), and the protocluster appears as a roughly mean-density region when viewed through the Lyman-$\alpha$ forest.
A search for protoclusters using only Lyman-$\alpha$ forest data (for example, \citealt{2015Tomopc,2016PCLya,cai:2016}) would probably miss protoclusters like this.

Hot bubbles are also seen in the 300-GadgetX thermal AGN run, but they do not extend into underdensities and the gas temperature in those regions remain cool ($\sim 10^4$ K). 
As a result, these regions exhibit some Ly$\alpha$ forest absorption such that the 300-GadgetX case is not as transparent as in 300-SIMBA. 

In the 300-GadgetMUSIC runs, high temperature gas is seen only in the direct vicinity of filaments within $\lesssim 1$Mpc from galaxies, and the Lyman-$\alpha$ transmission value is therefore the lowest among the different feedback prescriptions of The Three Hundred.
The Lyman-$\alpha$ absorption of this protocluster is clearly evident and would be easily detected in a search of the Ly-$\alpha$ forest.

\begin{figure*}
    \centering
    \scalebox{0.4}{\includegraphics{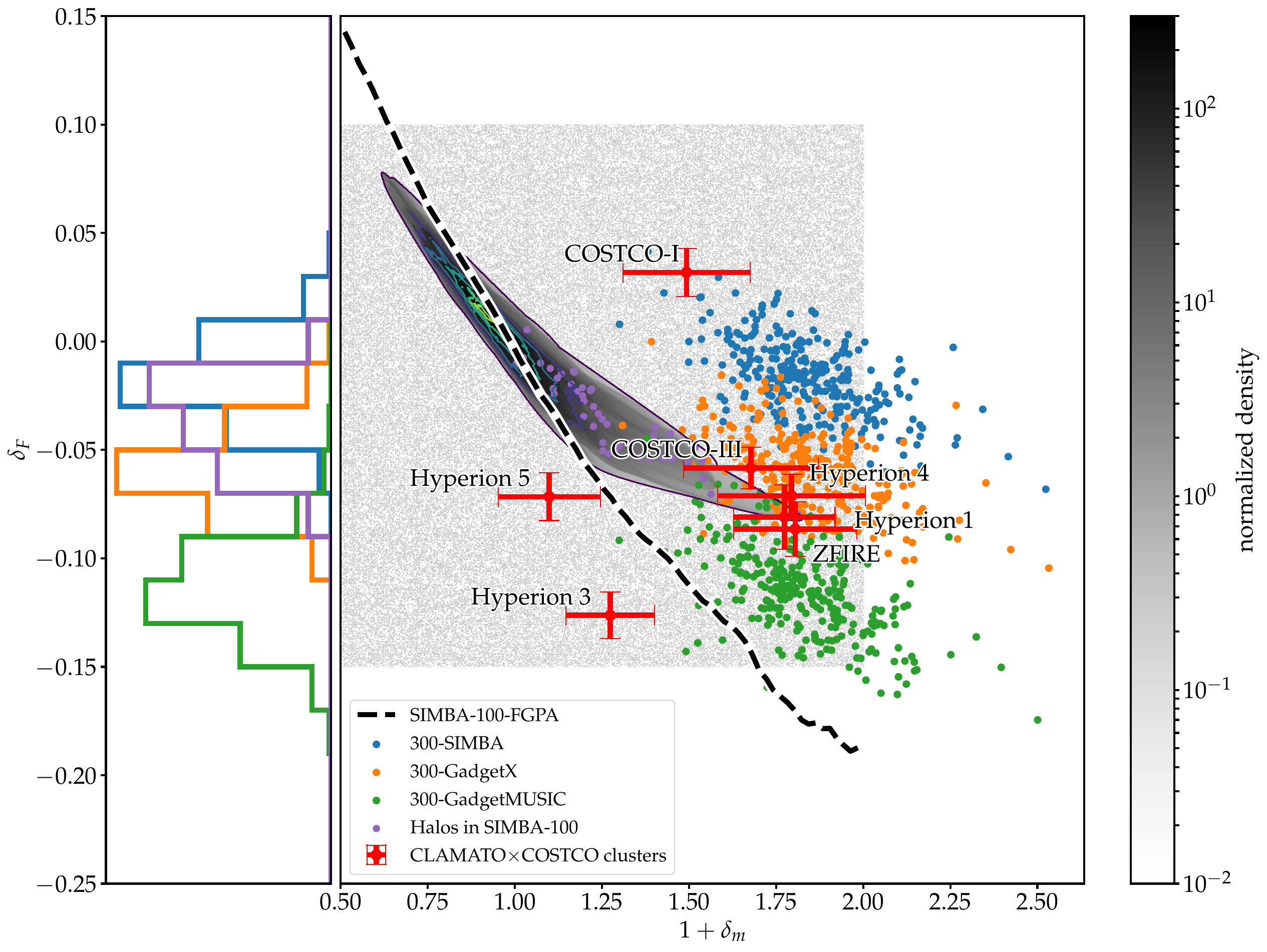}}
    \caption{The relation between \mG{} and \dfG{} in simulations as well as data. The scatter plots with blue, orange and green dots are measured from 300-SIMBA, 300-GadgetX, and 300-GadgetMUSIC, respectively. 
    The black dashed line depicts the FGPA relationship (equation \ref{eq:FGPA}) in SIMBA-100, where $\beta=1.6$.
    The shadow region and contours show the global relation of the two quantities in SIMBA-100 simulation box, where the purple points are the values of massive halos in SIMBA-100.
    The red points with error bars are the measurements from the \cxc{} protocluster. The left histogram shows the distribution of \dfG measured from massive protoclusters in our simulation samples, with the same color code as in the right panel. Most of the \cxc{} protoclusters appear to be deviating from the standard transmission-density relationship, while COSTCO-I shows a good match with the 300-SIMBA protoclusters that are implemented with jet feedback.}
    \label{fig:od}
\end{figure*}

Figure~\ref{fig:od} shows the smoothed Lyman-$\alpha$ forest transmission, \dfG{}, plotted against the correspondingly smoothed matter overdensity, \mG{}, measured for the ensemble of simulated protoclusters at $z=2.5$. 
We find a correlation between the two quantities within the individual simulation suites: \dfG{} generally decreases when a halo has higher \mG{}. 
The 300-SIMBA suite has the highest overall Lyman-$\alpha$ transmission among The Three Hundred runs, while the \dfG{} value of 300-GadgetX protoclusters shows an offset of $\Delta\dfG \sim -0.1$, i.e. more Lyman-$\alpha$ opacity. 
The 300-GadgetMUSIC suite exhibits the least transparent Lyman-$\alpha$ forest, which is probably due to the lack of AGN feedback in its feedback prescription.
For SIMBA-100, we show the matter-transmission relationship based on the fluctuating Gunn-Peterson relationship over the entire box as the black dashed line in Figure~\ref{fig:od}. 
In addition, we generate \texttt{pygad} Lyman-$\alpha$ skewers from the simulated HI gas density and show the skewer-based matter-transmission relationship over the entire box as a grey-scale contour.
The corresponding values centred on individual protoclusters within SIMBA-100 are also indicated separately as purple points.
The FGPA of SIMBA-100 appears to flatten out in \dfG{} with increasing \mG{}, but interestingly the points in 300-GadgetMUSIC seem to continue following the FGPA relation traced out by the lower-density SIMBA-100 points. 
This is in qualitative agreement with \citet{Viel:2013a}, who argued that stellar feedback does not have a significant effect on the power spectrum (i.e.\ fluctuations) of the Lyman-$\alpha$ forest.

The most massive halos in the SIMBA-100 box are generally not as massive as those in \sanbai{}, but they do not appear to follow a continuous relationship with those in 300-SIMBA. 
This is suspicious as one would expect the halos in SIMBA-100 to have similar properties as halos in 300-SIMBA due to the similar feedback prescriptions; we believe this to be due to the coarser resolution and the somewhat strengthened feedback 300-SIMBA, which we will discuss further in Section \ref{section:discussion}.
The data point for COSTCO-I closest to the trend of 300-SIMBA simulations: its overdensity is comparable with the less massive halos in The Three Hundred, while its transmission falls in the range of \dfG{} of 300-SIMBA.

We have also overplotted in Figure~\ref{fig:od} the measured \dfG{} from the CLAMATO Wiener-filtered maps and the \mG{} derived from the COSTCO constrained simulations, for the observed protoclusters listed in Table~\ref{tab:pc_info}.
The error bars were estimated from the ensemble of posterior realizations in the COSTCO simulation suite.

The \mG{} values of the observed \cxc{} protoclusters are generally $\mG>1$ as expected for overdensities, although those for Hyperion~3 and Hyperion~5 are only mildly overdense. 
This is not in contradiction with their identification as protoclusters, because protoclusters are often only mildly overdense over scales of multiple Megaparsecs \citep{2013PC_obs}.
Hyperion~3 and Hyperion~5 are components of the Hyperion proto-supercluster \citep{cucciati:2018}, which was shown by \citet{2022COSTCO} to be the progenitor of an eventual $L\sim 100$Mpc long filamentary supercluster with a total mass of $M_{h, z=0} \approx 2.5 \times 10^{15} \,h^{-1}\,M_\odot$. 
However, when compared with SIMBA simulation, these two protoclusters fall slightly below the FGPA \mG-\dfG{} relationship, i.e., they are more Lyman-$\alpha$ opaque than expected.
However, this discrepancy is at the $\sim 1-2\,\sigma$ level and is likely due to random scatter caused by observational uncertainties.

On the other hand, one sees clear deviations from SIMBA-100 FGPA in the other \cxc{} structures.
The majority of the protoclusters (COSTCO-III, ZFIRE, Hyperions 1 and 4) overlaps with the \mG{}-\dfG{} points from the 300-GadgetX cluster simulations, which incorporates the effect of thermal AGN feedback.
It indicates a clear departure from the global FGPA-like power-law relationship for these protoclusters, in that the amount of observed Lyman-$\alpha$ absorption is not commensurate with the matter overdensity traced by the COSMOS galaxies.

COSTCO-I exhibits an even more significant departure from the FGPA relationship, with a slightly positive \dfG{} value more usually indicative of underdensities. Its \mG-\dfG{} relationship does not appear consistent with those of the 300-GadgetX protoclusters, and lies among those of the 300-SIMBA simulations where AGN jet feedback has been implemented.

\section{Discussion} \label{section:discussion}
\subsection{Gravitational Collapse as the Origin of Protocluster Heating?} \label{subsection:nograv}
In this work, we made a comparison of the large-scale Lyman-$\alpha$ forest transmission of several galaxy protoclusters observed within the COSMOS field \citep{2022CLAMATO,2022COSTCO} with several hydrodynamical simulation suites that have varied galaxy and AGN feedback prescriptions. 
In particular, we aimed to investigate the anomalously low amount of absorption seen in the COSTCO-I protocluster at $z$=2.30 \citep{2023Preheat}.

Interestingly, the different feedback prescriptions lead to clearly distinct Lyman-$\alpha$ transmission values \dfG{} in the simulated protoclusters, 
especially if the controlling variable of underlying matter overdensity \mG{} is taken into account.
This indicates that measuring the FGPA-like relationship between Lyman-$\alpha$ flux and matter overdensity within galaxy protoclusters can be a sensitive probe of galaxy feedback. 
In a previous paper, \citet{2022FGPA} showed that hydrodynamical simulations with different feedback distributions exhibit different effective slopes in their global $\delta_F - \delta_m$ relationships, 
but this is a subtle effect that would probably require next-generation telescopes with $\sim 30$m diameter apertures to obtain sufficient signal to beat down the observational uncertainties. 
The results of Figure~\ref{fig:od}, on the other hand, 
suggests that focusing on galaxy protoclusters could provide clearer constraints on the feedback mechanisms.

Only two of the observed protoclusters within the \cxc{} volume appear to follow the global power-law FGPA relationship, while most of the protoclusters exhibit clear deviations from FGPA.
The majority of our protoclusters appear to follow the $\delta_F - \delta_m$ relationship traced by the 300-GadgetX simulations, which incorporate the effect of AGN thermal X-ray feedback.
The COSTCO-I protocluster, on the other hand, exhibits an even more extreme deviation from the FGPA, which places it within the relationship traced by the AGN jet-feedback model of the 300-SIMBA runs.  
Indeed, COSTCO-I is at a $\sim 5\sigma$ discrepancy with the simulated 300-GadgetMUSIC protoclusters, which model only stellar/supernova feedback as well as the generic hydrodynamical effect of gravitational shock-heating.

In \citet{2023Preheat}, we discussed the possibility that the transparent Lyman-$\alpha$ forest of COSTCO-I might simply be due to shock-heating of the cosmic web driven by nonlinear gravitational collapse within the overdense protocluster region.
Given that stellar feedback appears to have only a minor effect on the FGPA-like photo-ionisation equilibrium of the Lyman-$\alpha$ forest in the 300-GadgetMUSIC protoclusters (\citealt{Viel2013}),
the 300-GadgetMUSIC protoclusters set an upper limit to the impact of gravitational shock heating in protoclusters at this redshift. 
While there is a possibility that \sanbai{} has not fully converged for the Lyman-$\alpha$ forest (see Appendix \ref{ap:con_test}, also the discussion on simulation resolution in \citealt{Bourne:2015}), the clear deviations from FGPA in the majority of the observed protoclusters suggest that effects beyond simple gravitational collapse are likely at play. 
Upcoming datasets from \sanbai{} with higher resolutions (Cui et al., in prep.) would help to distinguish the effect of the unresolved baryon process in the case.
The fact also appears to be true for the COSTCO-III, Hyperion~1, Hyperion~4, and ZFIRE protoclusters: although their \dfG{} are not as high as COSTCO-I, they appear to deviate clearly from the adiabatic \mG-\dfG{} relationship.
Even for the most massive simulated protocluster progenitors ($M_{h, z=0} > 10^{15} M_\odot$) in the weak feedback model of 300-GadgetMUSIC, the Megaparsec-scale overdensities have not yet collapsed sufficiently at Cosmic Noon to drive enough gravitational shock heating to suppress the Lyman-$\alpha$ forest signal to the level seen in the majority of the \cxc{} protoclusters. 
This should also true for lower-mass, COSTCO-I-like halos $M_{h, z=0} =(4.6 \pm 2.2) \times 10^{14} \,h^{-1}\,M_\odot$, as they are even further from virialization.

In the literature, the importance of gravitational heating has mostly been discussed in the context of star formation quenching in galaxies. The pioneering work by \citet{2006Shock_quenching} suggests that the occurrence of gravitational shocks leads to the bimodality of galaxies, as the gravitational shocks shut down the continuous gas supply via cold streams. However, such a scenario is largely restricted by the redshift evolution of halo mass and shock heating mass (c.f., Figure 7 of \citealt{2006Shock_quenching}), limiting the effect of gravitational shocks to $z<2$ galaxies. 
More recent works, for example, \citet{2017Shock_quenching, 2017Shock_quenching_1}, also set the scope of such shock-heating-induced quenching to $z<2$.
In addition to the redshift restriction of gravitational-induced shocks, the shock radii of these $z\sim 2.5$ halos also suggest that the shock scenario is not preferable. 
As a rough estimation, the most massive halo in the $z=2.5$ snapshot of SIMBA-100 has a total mass of $8.53\times10^{13} M_\odot /h$, which translates into $r_{200c} = 946\,\mathrm{ckpc/h}$. 
The central galaxy group in COSTCO-I ($M_h \approx 6 \times 10^{13}\,h^{-1}\,M_\odot$; \citealt{2023Preheat}) has a similar halo mass, thus the extent of the central core halo spans a significantly smaller region than the observed scale of enhanced Lyman-$\alpha$ transparency in COSTCO-I, which is a few comoving Megaparsecs.
These scales would need to encompass the filamentary structures within the protocluster region, which are at lower densities than halos and therefore expected to collapse and shock-heat later at $z\sim 1$ (for example, \citealt{2019TNGgas}).

\subsection{Can AGN jet feedback explain COSTCO-I?}
Among the feedback prescriptions implemented in The Three Hundred, we find that the 300-SIMBA AGN jet feedback model yields the best match with the observed Lyman-$\alpha$ transmission of COSTCO-I. 
We also find a discrepancy between 300-SIMBA zoom-in runs and the SIMBA-100 cosmological volume, which ostensibly have similar feedback prescriptions. 

The massive cluster environment and the modified parameters from the original SIMBA model to 300-SIMBA contribute to the difference in transmission, but the simulation resolution might also be an issue.
In Appendix \ref{ap:con_test} we compare the SIMBA-100 run with two additional tests with the 300-SIMBA prescription. 
We find that the stronger jet feedback gives rise to the WHIM gas with $T \sim 10^7$ K by $z=2.5$, whereas the downgraded resolution changes the power-law $\rho-T$ relation in the diffuse cold gas regime. 
As stated in Section \ref{subsection:nograv}, the fact suggests that the 300-SIMBA runs have yet to reach resolution convergence, but this will not undermine the robustness of comparison between the different models implemented for \sanbai{}, nor does it affect our discovery that the majority of our observed protoclusters are in disagreement with FGPA.
Understanding the impact of resolution better necessitates zoom-in simulation suites with improved resolution (for example, TNG-Cluster; \citealt{2023TNG_cluster}).

We conclude that extreme transmission in the 300-SIMBA sample stems from the collective effect of a massive cluster environment, stronger feedback and low resolution, thus it is premature to claim that this particular AGN jet feedback model is necessarily an accurate depiction of the Lyman-$\alpha$ forest observed in COSTCO-I. 
Nevertheless, 300-SIMBA results are still useful for qualitative investigations into mechanisms that might facilitate large-scale heating of the IGM. In particular, it confirms the highly-collimated jet feedback can boost the large-scale heating of surrounding gas, and make massive protoclusters transparent in the Ly$\alpha$ forest. This is broadly consistent with results from previous works on the effect of the different feedback prescriptions in the SIMBA suite of simulations on the distribution and thermal state of gas in the IGM and CGM \citep{Bradley_2022, 2022Baryon_effect, 2024Baryon_partition, Yang_2024}.

Within the next few years, the IGM tomography component of the Subaru Prime Focus Spectrograph (PFS, \citealt{2014PFS, 2022PFS_GE}) will discover $\gtrsim 100$ galaxy protoclusters at $2.2<z<2.7$ to enable a statistical sample to compare with the hydrodynamical feedback models. 
By then, we aim to have resolved suites of simulations to compare with the observations.

\section{Summary and Conclusions}

In this paper, we examined the effect of various galaxy and AGN feedback mechanisms on the large-scale ($\sim$Mpc) Lyman-$\alpha$ forest transmission of galaxy protoclusters at Cosmic Noon ($z=2.5$).
This was motivated by the recent observational discovery of a $z=2.30$ galaxy protocluster, dubbed COSTCO-I \citep{2023Preheat}, 
which exhibits excess Lyman-$\alpha$ transmission (i.e.\ too little absorption) given the significant matter overdensity that it represents.

In addition to the $100\,\hMpc$ SIMBA-100 cosmological hydrodynamical simulation that features multiple stellar- and AGN-feedback mechanisms, we also analyzed a sample of cluster zoom-in simulations from \sanbai{} suite, 
in which each cluster was also simulated with three different feedback prescriptions (stellar feedback, AGN thermal feedback, and AGN jet feedback). 
We focused on the redshift snapshot at $z=2.50$ when most of the clusters were still in the protocluster stage. 
The simulated protocluster sample from \sanbai{} suite covers a range of cluster masses up to $3\times 10^{15} M_\odot/h$ at $z=0$, providing a good statistical sample to study COSTCO-I. 

The observable we focused on was the smoothed Lyman-$\alpha$ transmission, \dfG, as a function of the similarly-smoothed underlying matter overdensity \mG.
While these quantities can be straightforwardly calculated in the simulations, on the observational side we can do direct comparisons thanks to the combination of the 3D Lyman-$\alpha$ forest tomographic maps from the CLAMATO survey, 
as well as the COSTCO constrained simulations covering the same volume, that were based on the observed spectroscopic galaxy distribution in the COSMOS field.

We summarize our findings as follows:
\begin{itemize}
    \item When averaged over apertures of several Megaparsecs around individual galaxy protoclusters, we find clear differences in the Ly$\alpha$ transmission depending on the feedback model. 
    There is a weak dependence of Ly$\alpha$ transmission on the underlying matter density, but the different feedback models studied here all occupy distinct populations in the matter density-transmission ($\delta_m-\delta_F$) plane. This establishes that the Lyman-$\alpha$ forest around protoclusters is a potentially powerful probe of feedback mechanisms at Cosmic Noon ($z\sim 2-3$).
    \item From the $\delta_m-\delta_F$ relation for several of the observed \cxc{} protoclusters, we find deviations from the global power-law relationship that is expected to hold for the photo-ionised IGM at these epochs, i.e.\ there is a deficit of Lyman-$\alpha$ absorption in these structures. 
    These protoclusters are in marginal disagreement with the simulated protoclusters that incorporate stellar feedback only (300-GadgetMUSIC model). Since this model also sets an upper limit on the effect of shock heating from gravitational collapse, we argue that processes beyond gravitational shock-heating or stellar feedback are possibly at play in enhancing the transparency of the Ly$\alpha$ forest seen in these protoclusters.
    %Some form of AGN feedback appears to be necessary to explain this discrepancy.
    \item The COSTCO-I protocluster, in particular, exhibits an even stronger deviation ($\sim 5 \sigma$) from the FGPA relationship and 300-GadgetMUSIC. At face value, COSTCO-I is best matched to the protocluster population simulated with the AGN jet feedback model (300-SIMBA).
    However, lower-mass protoclusters in the SIMBA-100 cosmological hydrodynamical box, which nominally has a similar feedback model, do not form a continuous relation with the 300-SIMBA population in the $\delta_m-\delta_F$ plane.
    This is probably due to the large-scale extreme environment within the 300-SIMBA regions even in early times, but there is a possibility that \sanbai{} is not fully converged concerning the Lyman-$\alpha$ forest. Future analyses will need to be careful to ensure simulation convergence.
\end{itemize}

Strong AGN feedback in galaxy protoclusters has potential implications toward the use of the Lyman-$\alpha$ forest as a cosmological probe. 
Cosmological analyses of the Lyman-$\alpha$ forest power spectrum typically make the implicit assumption that the IGM at Cosmic Noon follows the FGPA, which stellar feedback largely does not perturb \citep{Viel:2013a}. 
Galaxy protoclusters represent regions of the Universe with the strongest density fluctuations at the $z\gtrsim 2$ epoch, therefore widespread AGN feedback in these structures could potentially smooth out the corresponding fluctuations that manifest into the Lyman-$\alpha$ forest.
While \citet{chabanier:2020} analyzed the effect of AGN jet feedback on the 1D Lyman-$\alpha$ forest power spectrum in the context of the HorizonAGN simulations, they did not check the specific case of galaxy protoclusters. 
It is therefore unclear whether their AGN model would reproduce the protocluster effects seen in this paper.
A recent paper by \citet{Rogers:2023} found an anomalous small-scale suppression in the Lyman-$\alpha$ forest power spectrum in tension with Planck, but our findings indicate that the effect of AGN feedback needs to be taken into account in such analyses.

In the near future, larger samples of Cosmic Noon galaxy protoclusters, such as those to be observed by the Subaru PFS Galaxy Evolution survey, in combination with converged hydrodynamical simulations, will provide strong constraints on the AGN feedback mechanisms at play in galaxy protoclusters.

\section*{Acknowledgements}

The authors thank the anonymous reviewer for the constructive feedback helping enhance the quality of this paper. The authors also thank Drew Newman for useful discussions on the correction of Wiener-filter effects in the observational data.
Kavli IPMU is supported by World Premier International Research Center Initiative (WPI), MEXT, Japan. This work was performed in part at the
Center for Data-Driven Discovery, Kavli IPMU (WPI).
This research was supported by Forefront Physics and Mathematics Program to Drive Transformation (FoPM), a World-leading Innovative Graduate Study (WINGS) Program, the University of Tokyo.
KGL acknowledges support from JSPS Kakenhi grant Nos. JP18H05868 and JP19K 14755.
WC is supported by the STFC AGP Grant ST/V000594/1, the Atracci\'{o}n de Talento Contract no. 2020-T1/TIC-19882 granted by the Comunidad de Madrid in Spain and the science research grants from the China Manned Space Project. He also thanks the Ministerio de Ciencia e Innovación (Spain) for financial support under Project grant PID2021-122603NB-C21, ERC: HORIZON-TMA-MSCA-SE for supporting the LACEGAL-III project with grant number 101086388, and the science research grants from the China Manned Space Project with No. CMS-CSST-2021-A01, and CMS-CSST-2021-A03.. 

This work has been made possible by the `The Three Hundred’ collaboration.\footnote{https://www.the300-project.org} The 300 simulations were performed at the MareNostrum Supercomputer of the BSC-CNS through The Red Española de Supercomputación grants (AECT-2022-3-0027, AECT-2023-1-0013), and at the DIaL -- DiRAC machines at the University of Leicester through the RAC15 grant: Seedcorn/ACTP317. For the purpose of open access, the author has applied a Creative Commons Attribution (CC BY) licence to any Author Accepted Manuscript version arising from this submission.

%%%%%%%%%%%%%%%%%%%%%%%%%%%%%%%%%%%%%%%%%%%%%%%%%%
\section*{Data Availability}

The results shown in this work use data from The Three Hundred galaxy clusters sample. These data are available on request following the guidelines of The Three Hundred collaboration, at \hyperlink{https://www.the300-project.org}{https://www.the300-project.org}. The data specifically shown in this paper will be shared upon request to the authors. The SIMBA-100 cosmological simulation is publicly available at \url{http://simba.roe.ac.uk/}.


\bibliographystyle{mnras}
\bibliography{example} % if your bibtex file is called example.bib

\appendix


\section{Correction for Observational Wiener-Filter Effects} \label{ap:obs_cali}
\begin{figure*}
    \centering
    \scalebox{0.35}{\includegraphics{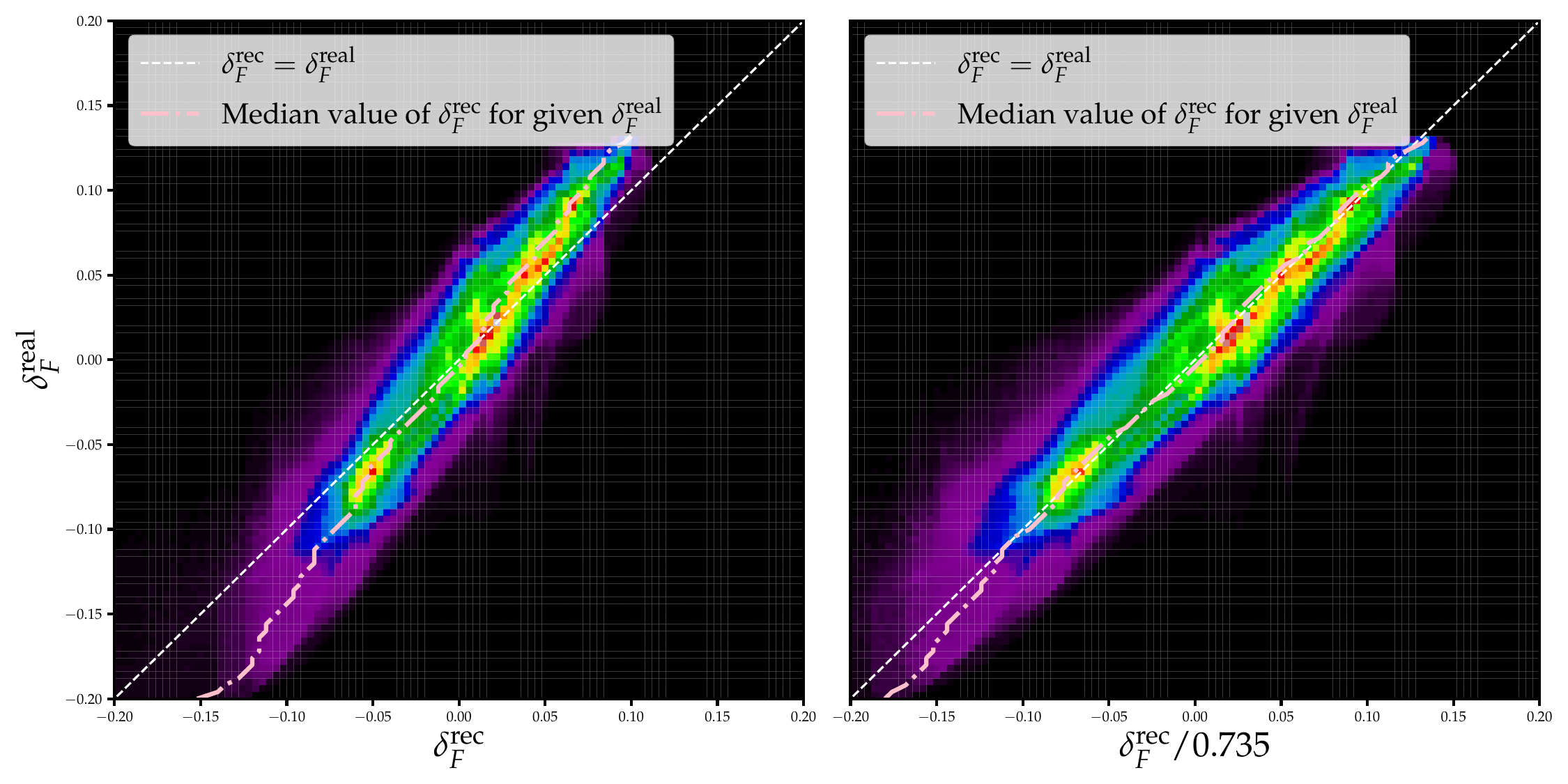}}
    \caption{Left: the relation between the Wiener filter-reconstructed flux $\delta_F^{\mathrm{rec}}$ and the underlying true value $\delta_F^{\mathrm{real}}$ in one COSTCO-FGPA realization, smoothed with a $7.5$ cMpc/h Gaussian kernel. Right: the same relation after applying a correction factor of $0.735$ to the reconstructed values. After the correction, $\delta_F^{\mathrm{rec}}$ and $\delta_F^{\mathrm{real}}$ are in good agreement.}
    \label{fig:recovery_test}
\end{figure*}

The Wiener filter --- which was applied to the CLAMATO Lyman-$\alpha$ forest absorption data to create 3D flux maps ---  has a prior of reconstructed transmission fluctuation $\delta_F^{\mathrm{rec}}=0$ (i.e.\ the mean transmission) for voxels in the reconstruction field. 
This introduces a bias toward the mean flux value in regions without sufficient line-of-sight sampling. 
According to Figure 7 of \citet{Lee:2014}, the Wiener filter will introduce a
tilt to the relationship between reconstructed flux $\delta_F^{\mathrm{rec}}$
and underlying smoothed matter density, in cases where the `true' flux $\delta_F^{\mathrm{true}}$ is known from numerical simulations. 
In this work, we correct this effect by fitting a relation between COSTCO-FGPA transmission data and the mock reconstruction result (see \citealt{2023Preheat}) based on $N$-body simulations, assuming the same sightline sampling as in CLAMATO. 
To account for the effect of smoothing in this work, we applied the same smoothing scale $7.5$ cMpc/h for the data. 
In Figure \ref{fig:recovery_test}, we find applying a correction factor of $0.735$ yields the best match between the reconstruction value and the underlying truth. 

\section{Calibration of Lyman-$\alpha$ Transmission at $z=2.5$ in Simulations} \label{ap:sim_cali}

According to \citet{Christiansen:2020}, for the optically thin case, $\tau \propto {1}/{\Gamma_{HI}}$ is a good approximation. 
Practically, this implies that changing the UV background is effectively multiplying the HI optical depth by a calibration factor.
This allows us to derive the factor $\Fuvb$ for each simulation and match the observation via rescaling the "raw" output of skewer generator.

For the SIMBA-100 simulation, we first obtained the "raw" Lyman-$\alpha$ optical depth $\tauraw$ grid in SIMBA-100 with \texttt{pygad}, and then derived a calibration factor $\Fuvb$ such that $\langle \Fuvb \cdot \tauraw\rangle = \langle F \rangle_{z=2.5, \mathrm{obs}}=0.79$, where the observational mean flux value is taken from \citep{Becker:2013}.

The Three Hundred simulations are centered on highly overdense regions that are not expected to yield the mean flux $\langle F \rangle$ when averaging over their Ly$\alpha$ sightlines.
Therefore, we run complementary cosmological simulations to carry out the \Fuvb{} calibration.
These complementary runs adopted the same cosmology, numerical resolution and feedback models as shown in \ref{tab:sims}. 
However, instead of zoom-in simulations of individual clusters, these complementary cosmological simulations are run with random initial conditions in a $(33\,\mathrm{cMpc/h})^3$ cosmological box. 
The choice of IC avoided the extremely massive protoclusters in \sanbai{} sample, thus providing a better calibration.

We present the calibration factors together with other information related to the UVB in Table \ref{tab:uvb_cali}. 

\begin{table}
    \centering
    \scalebox{0.8}{
    \centering
\begin{tabular}{ccccc}
Simulation & UVB & neutral Hydrogen abundance & ${\langle F \rangle}_{\mathrm{raw}}$ & \Fuvb \\ \hline
SIMBA-100 & HM12 & $1.48\times10^{-2}$ & 0.79 & 0.91 \\
300-GadgetMUSIC & HM01 & $5.97\times10^{-3}$ & 0.77 & 0.87 \\
300-GadgetX & HM96 & $3.14\times10^{-3}$ & 0.78 & 0.93 \\
300-SIMBA & HM12 & $4.20\times10^{-3}$ & 0.83 & 1.38
\end{tabular}}
    \caption{A summary table on the UVB adopted, the global neutral Hydrogen abundance, un-calibrated mean Lyman-$\alpha$ flux and calibration factor in each simulation. The calibration factor \Fuvb is the multiplier such that $\langle \exp(-\Fuvb \cdot \tauraw)\rangle = \langle F \rangle_{z=2.5, \mathrm{obs}}=0.79$, where $\tauraw$ is the optical depth before the calibration.}
    \label{tab:uvb_cali}
\end{table}

\section{Consistency test between SIMBA-100 and 300-SIMBA} \label{ap:con_test}
In this appendix, we show our observations on consistency between SIMBA-100 and 300-SIMBA. To achieve this, we run two test simulation boxes with randomly generated ICs (i.e., not in the environment of massive clusters) but keep the feedback prescription identical as in The Three Hundred. 
One simulation box has the identical resolution as SIMBA-100 and a box size of $25$ cMpc/h, while another box has a degraded resolution as in \sanbai{}, and a box size of $33$ cMpc/h. 

\begin{figure*}
    \centering
    \scalebox{0.3}{\includegraphics{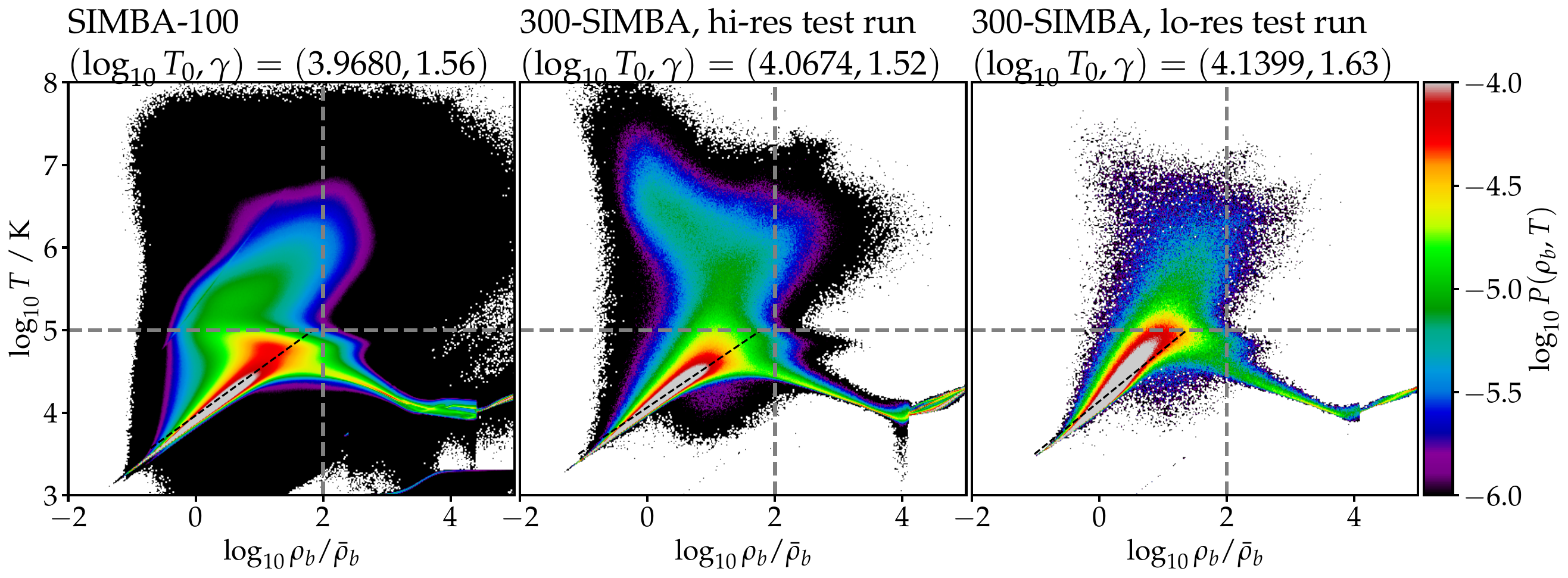}}
    \caption{From left to right: the mass-weighted $z=2.5$ temperature-density relation for SIMBA-100 and two (high resolution and low resolution) 300-SIMBA runs with ICs that are roughly at mean cosmic density. In each panel, the vertical grey dash line represents $\log_{10}\rho_b/\bar{\rho}_b = 2$ while the horizontal line is $T = 10^5 \mathrm{K}$. The two dash lines divide the $\rho-T$ phase space into diffused cold gas (lower left), warm-hot intergalactic medium (WHIM, upper left), hot halo gas (upper right) and condensed gas (lower right). In each panel, we add a reference dashed line to emphasize the linear $\rho-T$ relation of the diffuse cold gas regime and list the best-fit parameters, $\log_{10}T_0$ (in Kelvin) and $\gamma$ (defined in Section \ref{sec:intro}). All three panels show a similar pattern in the WHIM regime, while the low-resolution run has a distinct $\rho-T$ relation in the diffuse cold gas regime.}
    \label{fig:consistency_test}
\end{figure*}

Fig. \ref{fig:consistency_test} illustrates the mass-weighted temperature-density relation of gas in the simulation (TDR, e.g., \citet{2006MissBaryon},  \citet{2019TNGgas}) in SIMBA-100 and two test runs at $z=2.5$. 
For the connivance of discussion, we adopted the IGM phase classification of \citet{2015TDR}. The IGM gas is divided into four phases according to their temperature and density: diffuse cold gas (T $< 10^5 K, \rho_b / \bar\rho_b < 100$), warm-hot intergalactic medium (WHIM, T $> 10^5 K, \rho_b / \bar\rho_b < 100$), hot halo gas (T $> 10^5 K, \rho_b / \bar\rho_b > 100$) and condensed gas (T $< 10^5 K, \rho_b / \bar\rho_b > 100$). 
We find, in the diffuse regime, all the three simulation boxes reproduced a power-law $\rho$-T relation $T=T_0 (1+\delta)^{\gamma-1}$, but the low-resolution run has a larger power-law index $\gamma$, which may be relevant to the high overall Ly$\alpha$ transmission shown in Section \ref{section:result}. 
The pattern of gas distribution in the WHIM regime is different from simulation to simulation: most of WHIM in SIMBA-100 has $T\le 10^6$ K, though there is a small fraction of WHIM reaching a temperature of $10^7 - 10^8 $ K, which can be attributed to the high resolution and large box size of SIMBA-100. 
In the 300-SIMBA high-resolution test run, we find a branch of WHIM gas spanning from $(\log_{10}\rho_b/\bar\rho_b, \log_{10}T) = (1, 6)$ to $(\log_{10}\rho_b/\bar\rho_b, \log_{10}T) = (0, 7)$, implying the effect of stronger jet feedback in 300-SIMBA. The branch, however, only contributes a negligible portion of the total gas mass. 
A similar trend is found in the 300-SIMBA low-resolution test run, but the limitation on resolution makes the pattern less notable. 
We conclude that the modification of feedback strength mainly leads to a population of WHIM gas with $T > 10^7$ K at redshift $z=2.5$, and a lower resolution is likely to have an impact on the diffuse cold gas. 

\bsp	% typesetting comment
\label{lastpage}
\end{document}